        \documentstyle[12pt]{article}
\makeatletter
\def\address{\m@th\@ifnextchar[\@address{\@address[]}}

\def\@address[#1]#2{
\expandafter\def\expandafter\@addressname\expandafter
{\@addressname{
  \adr{#1}\ \parbox[t]{5in}{
     \ignorespaces #2}\par }}}
\def\@addressname{}
\def\adr#1{{\normalsize\unskip$^{#1}$}}
\def\@maketitle{%
\def\and{{\rm and}}
  \newpage
  \null
  {\centering
  \let \footnote \thanks
    {\Large\bf   \@title \par}%
    \vskip 1.5em%
      \lineskip .5em%
    {\bf\normalsize   \@author\par}
      \vspace{1em} 
    {\small \@addressname}
    
  }%
  \par
  \vskip 1.5em}
\def\section{\@startsection {section}{1}{\z@}{-3.5ex plus-1ex minus
    -.2ex}{1.5ex plus.2ex}{\reset@font\large\bf}}
\def\subsection{\@startsection{subsection}{2}{\z@}{-3.25ex plus-1ex
    minus-.2ex}{1.5ex plus.2ex}{\reset@font\normalsize\bf}}
\def\subsubsection{\@startsection
     {paragraph}{4}{\z@}{3.25ex plus1ex minus.2ex}{-1em}{\reset@font
     \normalsize\bf}}

\def\cite{\@ifnextchar[{\@tempswatrue\@citex}{\@tempswafalse\@citex[]}}

\def\@citex[#1]#2{%
\if@filesw\immediate\write\@auxout{\string\citation{#2}}\fi
\leavevmode\unskip\ \@cite{\@collapse{#2}}{#1}}

\def\@bylinecite{%
\@ifnextchar[{\@tempswatrue\@CITEX}{\@tempswafalse\@CITEX[]}%
}

\def\@CITEX[#1]#2{%
\if@filesw\immediate\write\@auxout{\string\citation{#2}}\fi
\leavevmode\unskip$^{\scriptstyle\@CITE{\@collapse{#2}}{#1}}$}

\def\@cite#1#2{[{#1\if@tempswa , #2\fi}]} %
\def\@CITE#1#2{{#1\if@tempswa , #2\fi}} %

\def\@collapse#1{%
{%
\let\@temp\relax
\@tempcntb\@MM
\def\@citea{}%
\@for \@citeb:=#1\do{%
\@ifundefined{b@\@citeb}%
{\@temp\@citea{\bf ?}%
\@tempcntb\@MM\let\@temp\relax
\@warning{Citation `\@citeb ' on page \thepage\space undefined}%
}%
{\@tempcnta\@tempcntb \advance\@tempcnta\@ne
\edef\MyTemp{\csname b@\@citeb\endcsname}%
\def\@tempa{\@temptokena=\bgroup}%
\afterassignment\@tempa %
\@tempcntb=0\MyTemp\relax}%
\ifnum\@tempcntb=0\relax%
\@tempcntb=\@MM
\@citea\MyTemp
\let\@temp = \relax
\else %
\edef\@tempd{\number\@tempcntb}%
\ifnum\@tempcnta=\@tempcntb %
\ifx\@temp\relax %
\edef\@temp{\@citea\@tempd}%
\else
\edef\@temp{\hbox{--}\@tempd}%
\fi
\else %
\@temp\@citea\@tempd
\let\@temp\relax
\fi
\fi
}%
\def\@citea{,}%
}%
\@temp %
}%
}%

\makeatother

\title{Emergence of Space-Time on the Planck Scale within the Scheme\\
of Dynamical Cellular Networks and Random Graphs}
        \author{Manfred Requardt}
        \address{Institut f\"ur Theoretische Physik,
         Universit\"at G\"ottingen,\\
         Bunsenstr. 9, 37073 G\"ottingen Germany}
\begin{document}
\maketitle

\section{Introduction}
The following is a very sketchy report about a research program, the
first steps of which have been accomplished in \cite{1} and
\cite{2}. Starting from the hypothesis that both the physics and the
mathematics of the Planck scale should be formed from intrinsically
discrete concepts, we develop a framework which is based on the
dynamical evolution of a class of '{\it cellular network
  models}', being capable (as we hope) of performing an '{\it unfolding phase
  transition}' from a presumed chaotic initial phase into  a new
phase which acts as an attractor in total phase space. This new phase
is assumed to carry a fine or '{\it super structure}' (kind of an
'{\it order parameter manifold}') which is then identified as the
discrete substratum underlying our ordinary continuous space-time (or
rather: the physical vacuum).

The building blocks of this emerging super structure are a particular
type of densely entangled subclusters of nodes/bonds of the underlying
network, which are conjectured to correspond to the '{\it physical
  points}' of the space-time continuum living on a much coarser scale;
thus reflecting the presumed internal complexity of '{\it space-time
  infinitesimals}'. Furthermore, these subclusters establish a certain
near- and far-order in the network, viz. a certain '{\it causal
  structure}', which is then again lifted to the ordinary space-time
continuum.

As a byproduct of our investigation we develop an arsenal of
mathematical tools (typically of a distinctly discrete flavor) which,
we think, will have a scientific value of their own well beyond the
field of Planck scale physics, as they are designed to analyze complex,
self-organizing and unfolding systems in general. On the other side,
they may help to create something like discrete analysis, topology,
geometry, dimension theory etc.

For further details concerning the various facets of our approach,
most notably the physical motivation and the physical/mathematical
sources, we refer the reader, due to lack of space, to the papers
\cite{1} and \cite{2}, the only exception being the extra mentioning
of the beautiful book of Bollobas (\cite{3}) about the theory of '{\it
  random graphs}', the results and concepts of which will play an
important role in the future analysis of the systems we are having in
mind. On the following pages we will mainly review the content of
paper \cite{2} which is primarily concerned with the analysis of the
'{\it unfolding phase transition}' in our cellular network models,
leading in the end to the emergence of a kind of (proto) space-time,
while in \cite{1} we were dealing in some detail with the development
of a version of discrete analysis and dimension theory and the like.

\section{The Dynamical Cellular Network Universe $QX$}
We now introduce one example of a cellular network law from a whole
class of possible laws (for more details see e.g. \cite{2}). The basic
constituents are its set of '{\it nodes}', $n_i$, and '{\it bonds}',
$b_{ik}$, connecting the nodes $n_i$ and $n_k$ with $i\neq k$, the labels 
$i,k$ running through some
countable index set. At each node $n_i$ sits an internal state $s_i$
taking its values (in our particular example!) in a certain subset of
the integers (modulo a possible ''unit charge'', say $q$, see below). 
The direct
interaction among the cells (represented by the nodes) is mediated by
certain bond states $J_{ik}$. In our example $J_{ik}\in \{+1,0,-1\}$,
where $J_{ik}=0$ means that the bond is dead or inactive at the time
(step) under discussion.

A '{\it local law}' is now introduced as follows: We assume that all
the nodes/bonds at '{\it (clock) time}' $t+\tau$, $\tau$ an elementary
clock time step, are updated according to a certain local rule which
relates for each given node $n_i$ and bond $b_{ik}$ their respective states at
time $t+\tau$ with the states of the nodes/bonds of a certain fixed
local neigborhood at time $t$.\\
Note: It is important that, generically, such a law does not lead
to a reversible time evolution, i.e. there will typically exist
attractors in total phase space (the overall configuration space of
the node and bond states).\\[0.5cm]
{\bf Remark}: A crucial ingredient of our network laws is what we
would like to call a '{\it hysteresis interval}'. We will assume that
our network starts from a densely entangled '{\it initial phase}'
$QX_0$, in which practically every pair of nodes is on average
connected by an '{\it active}' bond, i.e. $J_{ik}=\pm1$. Our dynamical
law will have a built-in mechanism which switches bonds off (more
properly: $J_{ik}=0$) if local fluctuations among the node states
become too large. There is then the hope that this mechanism may
trigger an '{\it unfolding phase transition}', starting from a local
seed of spontaneous large fluctuations towards a new phase (an
attractor) carrying a certain '{\it super structure}', which we would
like to relate to the hidden discrete substratum of space-time
(points).

One example of such a peculiar law is the following one:\\[0.3cm]
{\bf Local Law}: At each clock time step a certain '{\it quantum}' $q$
is transported between, say, the nodes $n_i,n_k$ so that 
\begin{equation}s_i(t+\tau)-s_i(t)=q\cdot\sum_k
  J_{ki}(t)\end{equation}
(i.e: if $J_{ki}=+1 $ a quantum $q$ flows from $n_k$ to $n_i$ etc.)\\
The second part of the law describes the back reaction on the bonds
(and is, typically, more subtle). This is the place where the socalled
'{\it hysteresis interval}' enters the stage. We assume two
'{\it critical parameters}' $0\leq\lambda_1\leq\lambda_2$ to exist
with:
\begin{equation}J_{ik}(t+\tau)=0\quad\mbox{if}\quad
  |s_i(t)-s_k(t)|=:|s_{ik}(t)|>\lambda_2\end{equation}
\begin{equation}J_{ik}(t+\tau)=\pm1\quad\mbox{if}\quad 0<\pm
  s_{ik}(t)<\lambda_1\end{equation}
with the special proviso that
\begin{equation}J_{ik}(t+\tau)=J_{ik}(t)\quad\mbox{if}\quad s_{ik}(t)=0
\end{equation}
On the other side
\begin{equation}J_{ik}(t+\tau)=\pm1\quad\mbox{or}\quad 0\quad\mbox{if}\quad
\lambda_1\leq\pm
  s_{ik}(t)\leq\lambda_2\end{equation}
provided that $J_{ik}(t)\neq 0$ or $J_{ik}(t)=0$.\\[0.3cm]
In other words, bonds are switched off if local spacial charge
fluctuations are too large and switched on again if they are too
small, their orientation following the sign of local charge
differences in between or remain inactive.\\[0.3cm]
{\bf Observation}: The above law fulfills two, in our view crucial,
properties: it consists of two parts, the one describing the effect of
'{\it geometry}' on '{\it matter}', the other the back effect of '{\it
  matter}' on '{\it geometry }' (in an up to now rather metaphorical
sense). It therefore may be viewed as a kind of '{\it proto gauge
  theory}'.\\[0.3cm]
{\bf Remarks}:i) The above dynamical law can or has to be supplemented
by appropriate boundary conditions.\\
ii) An important ingredient is the above '{\it hysteresis
 law}'. One
can speculate that, under appropriate conditions, it may be able to
catalyze an '{\it unfolding phase transition}' in the course of which
a substantial fraction of bonds becomes dead for a larger lapse of
time, thus inducing topological/geometrical changes in the primordial
network $QX$.\\
iii) Interrupting connections between nodes if local charge
differences become too large enhances the chance of '{\it pattern
  formation}' as the return to equilibrium is impeded.

\section{The Random Graph Aspect of the Network Dynamics}
We have to be very brief on this important and beautiful topic (for
more details see \cite{2}). As in the study of chaotic systems, where
one frequently goes over to a slightly simpler picture by employing
the socalled Poincar\'{e} map, it may be advantageous to concentrate
on the more ''graphical'' aspects of the underlying network dynamics,
viz. its topological/geometrical changes, in other words: its wiring
diagram.\\
(It goes without saying that, as in statistical mechanics, it
would be a fruitful idea to consider the network as a statistical
system and employ statistical concepts like '{\it entropy}',
'{\it correlation functions} etc., a route from which we have to
refrain due to lack of space; see however \cite{2}).

At each time step the underlying geometric structure may be considered
  as a '{\it
  graph}', $G(t)$, with '{\it node set}' $V$ and '{\it bond set}' $E$,
  $|V|,|E|$ their respective cardinalities. This picture becomes a
  dynamical one if we omit all the bonds $b_{ik}$ in $G(t)$ with
  $J_{ik}(t)=0$ and draw a bond $b_{ik}$ if $J_{ik}(t)\neq 0$. Our
  hypothesis is now that instead of following the full, complicated time
  evolution of the deterministically evolving network state, it is
  sensible to concentrate rather on the behavior of the corresponding
  time dependent graph $G(t)$, i.e. to focus attention on its wiring
  diagram and various typical '{\it graph observables}' as e.g. '{\it
  (average) degree of a node}', '{\it degree of connectedness}',
  structure  and number of certain '{\it subgraphs}', their mutual
  entanglement, viz. the '{\it geometrodynamics}' of the
  graph/network.

A, as we think, particularly beautiful tool (which was however
designed for an entirely different purpose in pure mathematics) is the '{\it
  random graph}' concept (see e.g. \cite{3}). As is the case
with the statistical method in general, this concept should be applied
to our network scenario with some care, since systems like ours are
possibly behaving only in a '{\it quasirandom}' manner, as their time
evolution follows a given deterministic law (which, however, may be
sufficiently '{\it ergodic}' or '{\it mixing}').

Our probabilistic model is now the following: With $n$ fixed vertices
given we make the set of possible graphs over the n-set $V$ a '{\it
  probability space}' by attributing each possible bond a probability 
$0\leq p\leq 1$. In other words: a given graph $G_m$ over
$V$ with $m$ bonds (an '{\it elementary event}') has the probability
\begin{equation}pr(G_m)=p^m\cdot q^{N-m}\quad\mbox{with}\quad N={n
    \choose 2},\; q=1-p \end{equation}
$N$ being the maximal possible number of bonds (simplex over $V$). On this
probability space one can now calculate all sorts of '{\it graph
  properties}' by introducing and employing the corresponding '{\it random
  functions}'. 

It was a fundamental observation of Erd\"os and Renyi
(for a proof of this deep result see e.g. \cite{4}) that many graph 
properties have a
socalled '{\it threshold function}' which is very reminiscent to '{\it
  phase transitions}' in statistical mechanics. To put it
sloppily:\\[0.3cm]
{\bf Observation (threshold functions)}: A large class of graph
properties have a threshold function $m(n)$, $m$ the number of edges,
$n$ the number of vertices, s.t. for $n\to\infty$ the graph under
discussion has the property $Q$ almost surely for $m>m(n)$ and almost
surely not for $m<m(n)$.
\section{The Unfolding Phase Transition}
With the help of the random graph concept, briefly described in section 3, 
we now want to sketch the scenario which, we hope, will lead to the 
emergence of a (proto)space-time structure (where, to be precise, nothing 
is said at the moment about the nature
 of ''classical'' time, which is postponed to forthcoming work. We rather 
concentrate on the emergence of the spacial aspect of the super structure 
under discussion).

We assume that the initial phase $QX_0$ behaves in some sense metastable 
(with respect to the local law introduced above) and that a sufficiently 
pronounced spontaneous fluctuation, leading to an avalanche of switched-off 
bonds, may drive the network $QX$
 towards another phase the nature of which we are going to describe now 
(for more details see \cite{2}).\\[0.3cm]
{\bf Conjecture}:i) Given that sufficiently many bonds are dead, a new 
superstructure may emerge in $QX$, the building blocks of which are 
conjectured to be (almost) maximal subsimplices $\subset QX$ or $G(t)$, 
called $mss$, i.e. maximal subclusters of no
des with (almost) every pair of nodes connected by an active bond.\\
ii) We want to relate these $mss$ with what we are used to call 
'{\it physical points}' in ordinary continuum physics, thus endowing them 
with a rich internal fine structure (which is largely hidden on the continuum 
level).\\
iii) The mutual entanglement of these $mss$, i.e. their fabric, is
assumed to underlie the causal structure we experience in continuum
space-time.\vspace{0.3cm}

The above $mss$ can be constructively generated for each given graph
$G(t)$. Starting from an arbitrary node $n_i$ one selects a node $n_j$
being connected with $n_i$, then a node $n_k$ being connected with
both $n_i$ and $n_j$ and so forth until the process terminates,
i.e. until there is no further node being connected with all(!) the
preceeding ones. Note however that at each intermediate step one can
possibly make different choices, leading to different $mss$ or not,
which, on the other side, will more or less overlap. This makes up the
fabric of the $mss$ and leads to a new type of '{\it hyper
  graph}'. With the help of the random graph approach one is now able
to analyze this complicated fabric. One is e.g. interested in the
typical size of the $mss$, their cardinality, their mutual overlap,
number of non-overlapping $mss$ in $G(t)$ etc.

\end{document}